# Simulation of Autonomous Industrial Vehicle Fleet Using Fuzzy Agents: Application to Task Allocation and Battery Charge Management

**Juliette Grosset**
IMT Atlantique
IRISA, UMR 6074

**Alain-Jérôme Fougères**
ECAM Louis de Broglie
IT Laboratory

**Ouzna Oukacha**
ECAM Louis de Broglie
IT Laboratory

**Moïse Djoko-Kouam**
ECAM Rennes – Louis de Broglie
IETR, UMR CNRS 6164

**Jean-Marie Bonnin**
IMT Atlantique
IRISA, UMR 6074

*Abstract: The research introduces a multi-agent simulation that uses fuzzy inference to investigate the work distribution and battery charging control of mobile baggage conveyor robots in an airport in a comprehensive manner. Thanks to a distributed system, this simulation approach provides high adaptability, adjusting to changes in conveyor agent availability, battery capacity, awareness of the activities of the conveyor fleet, and knowledge of the context of infrastructure resource availability. Dynamic factors, such as workload variations and communication between the conveyor agents and infrastructure are considered as heuristics, highlighting the importance of flexible and collaborative approaches in autonomous systems. The results highlight the effectiveness of adaptive fuzzy multi-agent models to optimize dynamic task allocation, adapt to the variation of baggage arrival flows, improve the overall operational efficiency of conveyor agents, and reduce their energy consumption.*

1. **INTRODUCTION**

The implementation of fleets of Autonomous Industrial Vehicles (AIV) in the context of Airport 4.0 presents a number of challenges, all of which are connected to the true degree of autonomy of these vehicles: employee acceptance, vehicle localization, traffic flow, failure detection, collision avoidance, and vehicle perception in dynamic environments. The different limitations and specifications developed by producers and potential consumers of these AIVs might be taken into consideration thanks to simulation.

Before starting to test AIV fleet traffic scenarios in often-complex airport situations, it is wise, if not essential, to simulate these scenarios (Hu and Zeigler, 2005). Furthermore, one of the primary benefits of employing simulations is that the outcomes can be utilized without requiring a scaling factor. The following are some key benefits of simulating AIV procedures (Tsolakis et al., 2019):
- reducing the time and cost of developing an AIV,
- minimizing potential operational risks associated with AIVs,
- allowing to assess the feasibility of different AIV circulation scenarios at a strategic or operational level,
- allowing a rapid understanding of the operations carried out by AIVs,
- identifying improvements in the layout configurations of the facilities hosting these AIVs.

Simulation also provides flexibility in terms of AIVs deployment and allows studying the sharing of responsibility between the central server and AIVs (local/global or centralized/decentralized balance) for the different operational decisions. Another advantage of simulations is the possibility to introduce humans into the scenarios in order to verify and validate, before the actual deployment of AIVs, the safety of the coexistence and possible interactions between these AIVs and human operators (Hentout, et al., 2019). Agent-based approaches are often proposed for the simulation of autonomous vehicles. They offer simulation contexts ranging from trajectory planning to optimal task allocation, while allowing collision and obstacle avoidance (Jing, et al., 2020).

Our current research focuses on the use of fuzzy agents to handle the levels of imprecision and uncertainty involved in modeling the behavior of simulated vehicles (Fougères, 2013). Indeed, fuzzy set theory is well suited to the processing of uncertain or imprecise information that must lead to decision-making by autonomous agents, used in activities such as the simulation of AIVs in an airport or product design (Fougères and Ostrosi, 2013).

Fuzzy agents can track the evolution of fuzzy information from their environment and from agents (Ghasem-Aghaee and Ören, 2003). By interpreting the fuzzy information they receive or perceive, fuzzy agents interact within the multi-agent system of which they are a part. For example, a fuzzy agent can discriminate a fuzzy interaction value to assess its degree of affinity (or interest) with another fuzzy agent (Ostrosi, et al., 2012).

Therefore, we conduct a thorough study into the use of fuzzy inference in multi-agent simulations to maximize work distribution and battery management for airport mobile baggage conveyor robots. The suggested simulation method is made to be flexible, accounting for dynamic elements like changes in workload, battery capacities, and agent-to-infrastructure communication. The findings show that an adaptive fuzzy multi-agent model can lower energy usage, adjust to changes in baggage arrival flows, and greatly increase operational efficiency.

The structure of this article is as follows. We start by introducing a state-of-the-art in work allocation using fuzzy agents. Next, we compare five different types of techniques in a case study on fuzzy work allocation. We offer three enhancements utilizing fuzzy heuristics in Section 4. Following our conclusion on the suggested fuzzy dynamic task allocation systems, we discuss various perspectives.

## 2. FUZZY AGENT-BASED TASK ALLOCATION SIMULATION

### 2.1. Task Allocation

Task Allocation (TA) consists of optimally assigning a set of tasks to be performed by agents, actors, robots or processes, grouped and organized within a cohesive system. This is the case for mobile multi-robot systems, AIV fleets, and applications in airports (El-Ansary, et al., 2017).

In the field of mobile robotics, a taxonomy has been defined to better characterize allocation and assignment functions to robots (De Ryck, et al., 2020): Single Task for a Single Robot (STSR), Multiple Tasks for a Single Robot (MTSR), and Multiple Tasks for Multiple Robots (MTMR). These classifications enable tasks to be assigned to one or multiple robots, with various tasks being allocated to heterogeneous or multitasking robots. Moreover, two types of allocation modalities can be defined: instantaneous allocation and allocation extended in time. This last is linked to synchronization and precedent or time window constraints. As many combinations as exhaustively detailed by numerous surveys on the issue of multi-robot TA.

A variety of models have been proposed for TA solutions, including those based on optimization, such as exact algorithms, dynamic programming, and (meta-)heuristics (Khamis et al., 2015); the Contract Net Protocol, which states that in an agent-based system, an initiating agent calls for proposals from all agents, selects the best proposal, and then notifies all agents (Karur et al., 2021); and the market concept, which includes an auctioneer's announcement, bidders' submissions, the auctioneer's selection, and the auctioneer's award (Hussein and Khamis, 2013).

Furthermore, different types of optimization objectives can be defined for this task allocation (De Ryck, et al., 2020): cost objectives (travel costs, such as time, distance or fuel consumption), behavior objectives (ability of a robot to perform a task), reward objectives (payoff for performing a task), priority objectives (urgency to perform a task), and utility objectives (subtracting the cost from the reward or fitness).

Task allocation and planning are often managed centrally, even semi-centrally when global and local planning are differentiated (Mariani, et al., 2021). For the proper functioning of autonomous and dynamic systems, the requirements of flexibility, robustness and scalability, lead to consider decentralized mechanisms to react to unexpected situations. Autonomy and decentralization are two excessively linked notions to the extent that an autonomous system operates and makes decisions autonomously (De Paula, et al., 2020). The problem of task allocation can also be thought of in a decentralized way (Grosset, et al., 2024).

For reasons of flexibility, robustness and scalability necessary in an Industry 4.0 or Airport 4.0 context, we are interested in decentralized task allocation solutions. These solutions, decomposed below, must be able to assign tasks to a fleet of robots. Particularly, solutions based on the market concept can easily be applied in a distributed context, where each mobile robot can become an auctioneer (Daoud, et al., 2021). For each situation, a single mobile robot is appointed auctioneer, and retains this role until the situation is definitively managed

### 2.2. Fuzzy Agent-Based Simulation

Many agent-based approaches are proposed for the simulation of autonomous vehicles (Huang, et al., 2022). They offer simulation contexts ranging from trajectory planning (Kou, et al., 2019) to optimal task allocation, while allowing collision and obstacle avoidance (Grosset, et al., 2023). Fuzzy set theory is well suited to the processing of uncertain or imprecise information that must lead to decision-making by autonomous agents. Therefore, our current research focuses on the use of fuzzy agents to handle the levels of imprecision and uncertainty involved in modeling the behavior of simulated vehicles (Grosset, et al., 2024).

Most of the control tasks performed by autonomous mobile robots have been the subject of performance improvement studies using fuzzy logic: navigation (Yerubandi, et al., 2015), obstacle avoidance (Meylani, et al., 2018), path planning (Patle, et al., 2019), motion planning (Nasrinahar and Chuah, 2018), localization of mobile robots (Alakhras, et al., 2020), and intelligent management of energy consumption (Lee and Nugroho, 2022).

An agent-based system is fuzzy if its agents have fuzzy behaviors or if the knowledge they use is fuzzy (Ostrosi, et al., 2012). This means that agents can have: 1) fuzzy knowledge (fuzzy decision rules, fuzzy linguistic variables, and fuzzy linguistic values); 2) fuzzy behaviors (the behaviors adopted by agents because of fuzzy inferences); and 3) fuzzy interactions, organizations, or roles.

### 3. APPLICATION: TASK ALLOCATION AND BATTERY CHARGE MANAGEMENT SIMULATION

This case study proposes the simulation of mobile robots conveying baggage fleet in an airport (we will keep the name "AIV" for these conveyors). Figure 1 shows the simulator's HMI, which allows on the one hand, to visualize the arrival of baggage and the movements of 5 AIVs, and on the other hand, to follow the evolution of the four levels of indicators of the simulation: 1) energy level, 2) baggage level, 3) charge level, and 4) time level.

**FIGURE 1**
**HMI OF THE SIMULATION APPLICATION AND ORIENTED GRAPH MODELLING THE CIRCULATION OF AIVS (DISTANCES IN METERS)**

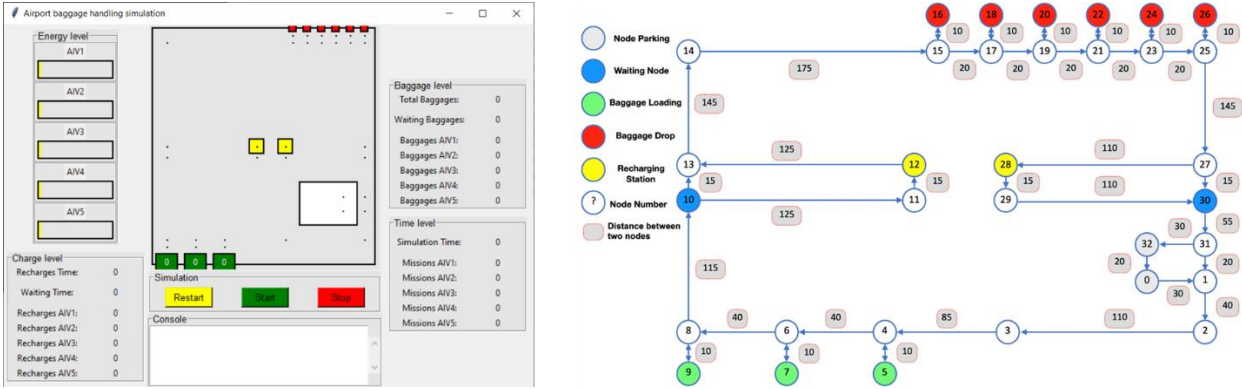

Effective management of these AIVs requires an integrative approach that considers several factors, including the baggage arrival flow, the operational availability of the AIVs, their energy consumption, their communication, among themselves and with the infrastructure, and their adaptation to changing environmental conditions. In the case study, we analyze the TA performed by a supervisor who questions AIVs to know their task completion costs. Through 8 scenarios, we will progressively introduce fuzzy inferences to determine the costs of task completion, battery recharging and speed adjustment.

### 3.1. The Simulation Framework
In previous work, we have developed a fuzzy agent model to test dynamic strategies for AIVs in simulation (Grosset, et al., 2024). The objective is to have a fuzzy agent-based modeling and simulation system designed generically to test different scenarios, but also different types of circulation plans. An infrastructure is deployed in the environment. It is composed of a circulation plan and active elements, such as beacons, tags, the two charging stations and the two types of treadmill for baggage entry and exit. Static or dynamic obstacles (e.g., operators) may be present in the environment.

AIV fuzzy agents perform missions defined by paths on the traffic plan. AIV fuzzy agents are equipped with a radar to avoid collisions and have knowledge about the environment and other agents to operate and cooperate. AIV fuzzy agents communicate with each other with different types of standardized messages. AIV fuzzy agents have fuzzy and uncertain knowledge, but also incomplete and fragmented, in order to adapt to situations that are themselves uncertain. Baggage are objects managed by the environment: arrival flow on the entry treadmill, tracking of its localization, and exit from the circuit on the exit treadmill.

### 3.2. Task Allocation with Basic Strategies
In this section, we provide a comparative analysis of three basic types of auction-based task allocation strategies: random TA, FIFO TA, and AIV availability-based TA. Each of these strategies is tested in a scenario:
- *Sc1* (Random) is a TA scenario where missions are assigned to the AIV agents only randomly.
- *Sc2* (FIFO) is a TA scenario where missions are assigned to AIV agents using a queuing mechanism.
- *Sc3* (Available) is a TA scenario where missions are assigned to the most available AIV agents.

We simulated these three scenarios for 100 bags. We seek to minimize the maximum number of pending bags at a given time, the total simulation time, the average time to complete a mission per AIV agent, the number of missions completed per AIV agent during the simulation, and the activity rate per AIV agent. The simulation results are presented in Table 1.

*Random Strategy*

The maximum number of pending bags is high, the simulation time is also high, and the allocation of missions and the activity rates of AIV agents are poorly balanced (the average activity rate at 0.72 is low). The random strategy does not allow allocation to AIV agents that are a priori available, which very quickly leads to pending bags to be processed and therefore poor results.

*FIFO Strategy*

This strategy brings a clear improvement in the results. The maximum number of pending bags is very low, the simulation time is very correct, the allocation is almost uniform (only the stops for recharging the batteries cause imbalances), and the occupancy rate of the AIV agents is much better (0.84).

*Available Strategy*

This strategy produces the best results, except for the maximum number of pending bags. Allocating a mission to an AIV agent that is more available than the others are therefore improves the results. However, it is necessary to better manage the allocation based on pending bags and energy consumption to consolidate (or even optimize) this strategy.

**TABLE 1**
**TASK ALLOCATION SIMULATION RESULTS IN SCENARIOS SC1, SC2 AND SC3**
**(100 BAGS)**

| Scenarios | Random | FIFO | Available |
|---|---|---|---|
| Maximum nb of pending bags | 19 | 4 | 8 |
| Simulation time | 2270s | 1942s | 1846s |
| Average mission time per AIV (in s) | [81,81,83,83,81] | [80,82,83,81,83] | [81,80,81,83,81] |
| Nb of missions completed by AIV | [26,26,14,14,20] | [21,21,19,21,18] | [22,21,20,19,18] |
| Work rate per AIV | [0.93,0.93,0.51,0.51,0.71] | [0.87,0.89,0.81,0.88,0.77] | [0.97,0.91,0.88,0.85,0.79] |

### 3.3. Task Allocation with Fuzzy Strategies

Fuzzy logic allows bettering understanding natural, uncertain, imprecise and difficult to model phenomena by relying on the definition of *if-then fuzzy* rules and membership functions (linguistic variables) to *fuzzy sets* (Zadeh, 1965). Therefore, in this section, we propose an analysis of task allocation by auction based on a fuzzy inference approach. Two scenarios are studied.

The first scenario, *Sc4*, implements a TA strategy in which each AIV agent uses a fuzzy model to determine the cost of handling a mission (picking up and dropping off a baggage). Three linguistic input variables, $I_1$ to $I_3$, and one linguistic output variable, $I_3$, are used (Figure 2):
- $I_1$: Availability, which represents "the availability of the AIV agent",
- $I_2$: DistanceTaget, which represents "the distance from the baggage drop-off location",
- $I_3$: EnergieLevel, which represents "the energy level of the AIV agent",
- $O_1$: Cost, which represents "the cost of handling a mission".

The second scenario, *Sc5*, takes the strategy of *Sc4* and adds energy management with a second fuzzy model. With this new fuzzy model, the AIV agents determine whether they will need to recharge during a mission, which allows them to refine calculation of the mission cost. The six linguistic variables, $I_1$ to $I_4$ and $O_1$ to $O_2$, used in this scenario are (Figure 3):

- $I_1$: Availability, which represents "the availability of the AIV agent",
- $I_2$: DistanceTaget, which represents "the distance from the baggage drop-off location",
- $I_3$: EnergieLevel, which represents "the energy level of the AIV agent",
- $I_4$: DistanceStation, which represents "the distances to the 2 recharging stations",
- $O_1$: Cost, which represents "the cost of handling a mission",
- $O_2$: Recharge, which represents "the decision to recharge the battery or continue a mission".

*Fuzzy Strategy in Sc4*

The results with this new strategy are generally good (Table 2): low maximum number of pending bags, good overall simulation time, good distribution of missions between AIV agents and good average AIV activity rate (0.88). However, few elements of uncertainty are considered (three linguistic variables at the input and one at the output). The introduction of other fuzzy elements (nuances in the simulation parameters) should improve the results, particularly in terms of maximum number of pending bags and management of battery recharges.

*Fuzzy Strategies in Sc5*

In this new scenario, the raw results of the TA (Table 2) are slightly worse than in *Sc4*: same maximum number of pending bags, slightly longer overall simulation time, worse distribution of missions between AIV agents and worse average AIV occupancy rate (0.82). However, the overall recharge time is lower in this scenario (Table 3), which can allow a greater availability of AIV agents (an area of improvement for the next scenarios).

**FIGURE 2**
**SCENARIO *SC4*: FUZZY MODEL USED AND HMI VIEW AT THE END OF THE SIMULATION**

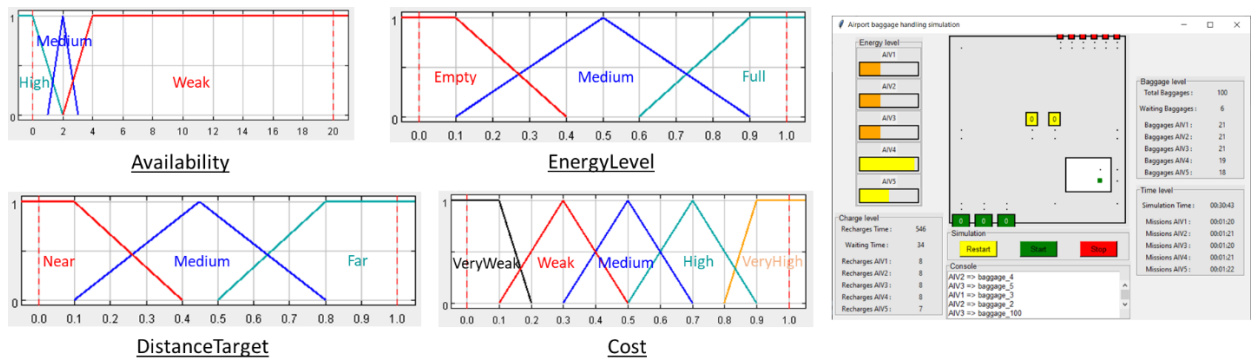

# FIGURE 3
## SCENARIO *SC5*: FUZZY MODEL USED AND HMI VIEW AT THE END OF THE SIMULATION

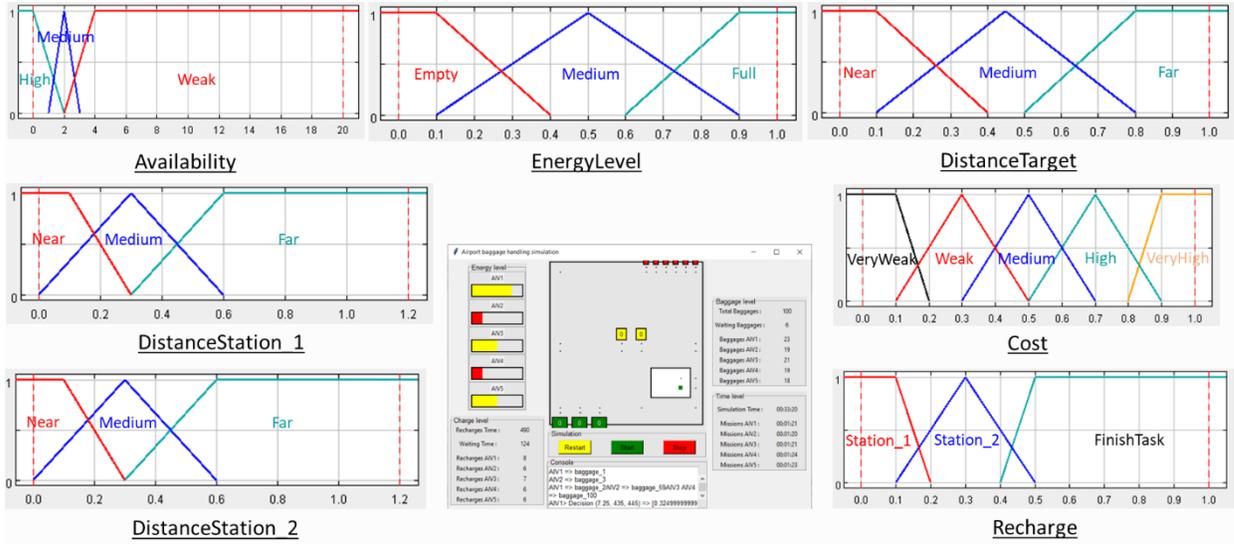

# TABLE 2
## TASK ALLOCATION SIMULATION RESULTS IN SCENARIOS SC4 AND SC5 (100 BAGS)

| Scenarios | Sc4 | Sc5 |
|---|---|---|
| Maximum nb of pending bags | 6 | 6 |
| Simulation time | 1843s | 2000s |
| Average mission time per AIV (in s) | [80, 81, 80, 81, 82] | [81, 80, 81, 84, 83] |
| Nb of missions completed by AIV | [21, 21, 21, 19, 18] | [23, 19, 21, 19, 18] |
| Work rate per AIV | [0.91,0.92,0.91,0.84,0.80] | [0.93,0.76,0.85,0.80,0.75] |

# TABLE 3
## RECHARGE SIMULATION RESULTS IN SCENARIOS SC4 AND SC5 (100 BAGS)

| Scenarios | Sc4 | Sc5 |
|---|---|---|
| Recharge time | 546s | 490s |
| Waiting time for recharges | 34s | 16s |
| Nb of recharges | 39 | 33 |
| Distribution of nb of recharges per AIV | [8, 8, 8, 8, 7] | [8, 6, 7, 6, 6] |

## 4. IMPROVEMENT USING FUZZY HEURISTICS

Now, we propose to increase the relevance of previous auction TA scenarios based on a fuzzy inference approach, by integrating other types of realistic constraints concerning battery recharging and AIV agent speed adjustment made possible by a stronger knowledge of the fleet traffic and mission management context (increased awareness). Three scenarios are studied (*Sc6*, *Sc7* and *Sc8*) to show that specific heuristics allow us to treat certain situations quite finely and to increase the collective/global performances of the AIV agents. The results are presented in Table 4 for task allocation and Table 5 for battery recharging.

$Sc6$ consists of completing scenario $Sc5$ to determine in which station the AIV agents can recharge in order to minimize the waiting times for recharging, based on knowledge of the context of occupation of the stations and the needs of the other AIV agents (therefore more awareness for the agents). The seven linguistic variables, $I_1$ to $I_5$ and $O_1$ to $O_2$, used in this scenario are (Figure 4):

- $I_1$: Availability, which represents "the availability of the AIV agent",
- $I_2$: DistanceTaget, which represents "the distance from the baggage drop-off location",
- $I_3$: EnergieLevel, which represents "the energy level of the AIV agent",
- $I_4$: DistanceStation, which represents "the distances of the 2 recharging stations",
- $I_5$: AvailabilityStation, which represents "the availability of the recharging stations",
- $O_1$: Cost, which represents "the cost of handling a mission",
- $O_2$: Recharge, which represents "the decision to recharge the battery or continue a mission".

$Sc7$ takes up the strategy of $Sc6$ and adapts the recharging rate (80 or 100%) in order to increase their availability if the flow of incoming baggage increases and therefore if the number of pending bags is likely to increase. The seven linguistic variables, $I_1$ to $I_5$ and $O_1$ to $O_2$, used in this scenario are (Figure 5):

- $I_1$: Availability, which represents "the availability of the AIV agent",
- $I_2$: DistanceTaget, which represents "the distance from the baggage drop-off location",
- $I_3$: EnergieLevel, which represents "the energy level of the AIV agent",
- $I_4$: DistanceStation, which represents "the distances to the 2 recharging stations",
- $I_5$: AvailabilityStation, which represents "the availability of the recharging stations",
- $O_1$: Cost, which represents "the cost of handling a mission",
- $O_2$: RechargeRate, which represents "the variable energy charge rate" (80 or 100%).

$Sc8$ consists of increasing $Sc7$ by adapting/regulating the speed of the AIV agents according to the flow of baggage arrivals and therefore the potential increase in the number of pending bags to be processed, but also according to the speed, the proximity of other AIV agents (use of observed and safety distances). The eight linguistic variables, $I_1$ to $I_6$ and $O_1$ to $O_2$, used in this scenario are (Figure 6):

- $I_1$: Availability, which represents "the availability of the AIV agent",
- $I_2$: DistanceTaget, which represents "the distance from the baggage drop-off location",
- $I_3$: EnergieLevel, which represents "the energy level of the AIV agent",
- $I_4$: DistanceStation, which represents "the distances to the 2 recharging stations",
- $I_5$: AvailabilityStation, which represents "the availability of the recharging stations",
- $I_7$: Urgency, which represents "the urgency in relation to the number of pending bags",
- $O_1$: Cost, which represents "the cost of handling a mission",
- $O_2$: RechargeRate, which represents "the variable energy charge rate" (80 or 100%).

*Results of Fuzzy Inferences in Sc6*

This is the implementation of a first heuristic to improve the TA but also the recharge decision. The objective is to minimize the waiting time for a recharge when an AIV agent must be available to take baggage. The results for TA are slightly better than in $Sc5$ (Table 4): the same maximum number of pending bags, a slightly shorter overall simulation time, a rather homogeneous average mission completion time, a better distribution of missions between AIV agents, and an average AIV activity rate that is roughly the same (0.82). However, if the overall recharge time is the same (Table 5), the waiting time for recharges is significantly lower (14s).

*Results of Fuzzy Inferences in Sc7*

Second heuristic proposed in order to increase the availability of AIV agents so that they can take baggage according to their arrival flow while minimizing the waiting time for their recharges. In this scenario, the results for TA are significantly better than in the $Sc6$ scenario (Table 4): the same maximum number of pending bags, but a shorter overall simulation time, a more homogeneous average mission completion time, a better distribution of missions between AIV agents and a higher average AIV activity rate (0.84). Regarding battery recharges, the results are of the same order for both scenarios (Table 5): an identical overall recharge time, with in *Sc7*, a slightly higher waiting time for recharges (18s).

*Results of Fuzzy Inferences in Sc8*

A third heuristic was proposed in order to adjust speed of the AIV agents to minimize the maximum number of pending bags when the flow of baggage arrivals increases. The results for TA are much better than in *Sc7* (Table 4): the same maximum number of pending bags, but a much lower overall simulation time (a consequence of the adaptation of speeds of AIV agents when necessary), an average time of completion of the missions and a distribution of the missions between the AIV agents always homogeneous, and finally, a lower average occupancy rate of the AIV agents (0.79), because the last two AIV agents are less requested due to the adaptation of the speeds of the first 3, in particular their increase in speed to respond to the increase in the flow of baggage arrivals. As for the battery recharges, the results are less good (Table 5): the increase in the speeds of the AIV agents has an energy cost!

**TABLE 4**
**TASK ALLOCATION SIMULATION RESULTS IN SCENARIOS SC6; SC7 AND SC8, FOR 100 BAGS**

| Scenarios | Sc6 | Sc7 | Sc8 |
|---|---|---|---|
| Maximum nb of pending bags | 6 | 6 | 6 |
| Simulation time | 1964s | 1896s | 1675s |
| Average mission time per AIV (in s) | [79,79,80,80,81] | [79,80,80,80,80] | [67,65,67,65,67] |
| Nb of missions completed by AIV | [22,22,20,16,20] | [22,22,21,18,17] | [22,22,22,19,15] |
| Work rate per AIV | [0.88, 0.88, 0.81, 0.65, 0.82] | [0.92, 0.93, 0.89, 0.76, 0.72] | [0.88, 0.85, 0.88, 0.74, 0.6] |

**TABLE 5**
**RECHARGE SIMULATION RESULTS IN SCENARIOS SC6, SC7 AND SC8, FOR 100 BAGS**

| Scenarios | Sc6 | Sc7 | Sc8 |
|---|---|---|---|
| Recharge time | 490 | 490 | 736 |
| Wait time for recharges | 14 | 18 | 119 |
| Nb of recharges | 33 | 33 | 49 |
| Distribution of nb of recharges per AIV | [7, 7, 7, 5, 7] | [7, 7, 7, 6, 6] | [11, 11, 11, 9, 7] |

**FIGURE 4**
**SCENARIO *SC6*: FUZZY MODEL USED AND HMI VIEW AT THE END OF THE SIMULATION**

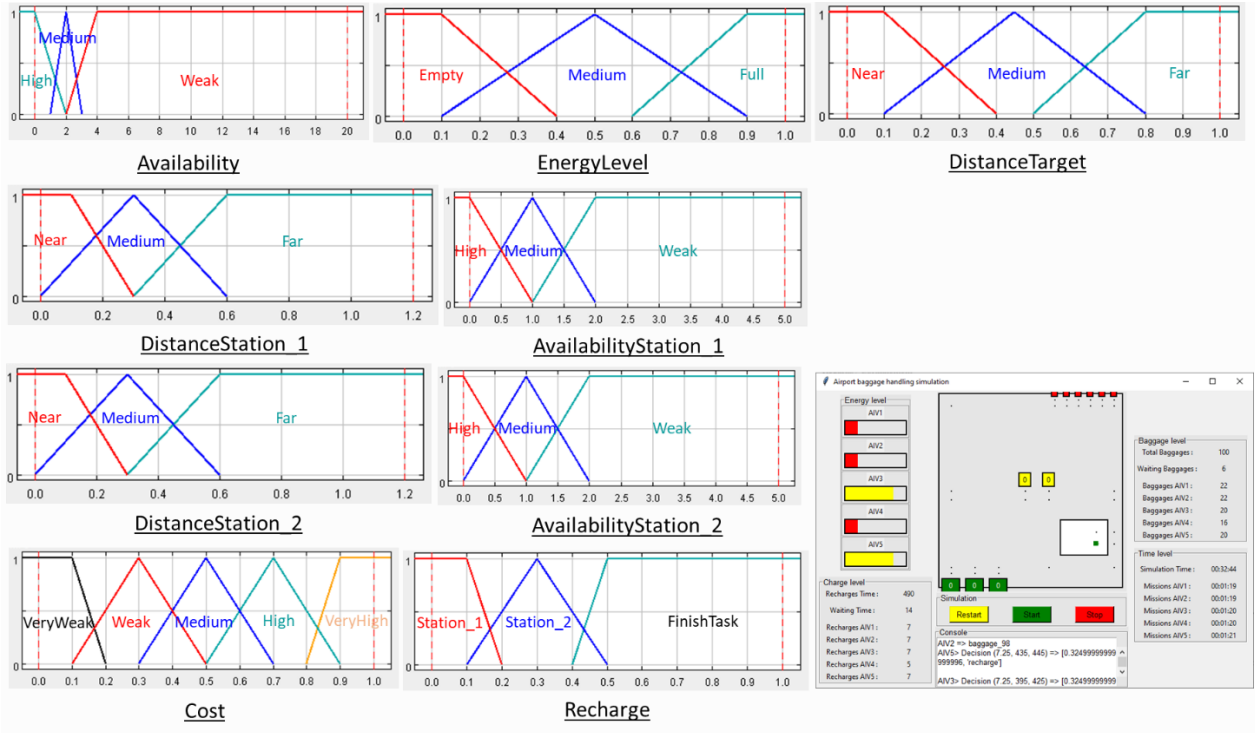

**FIGURE 5**
**SCENARIO *SC7*: FUZZY MODEL USED AND HMI VIEW AT THE END OF THE SIMULATION**

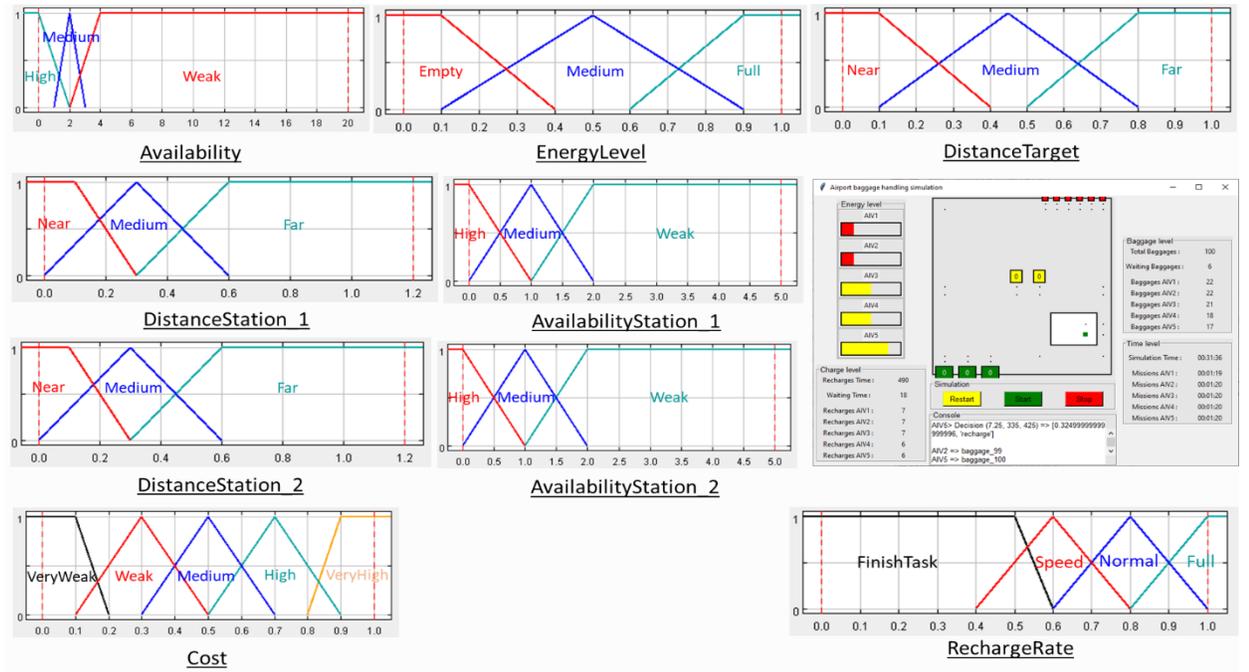

**FIGURE 6**
**SCENARIO *SC8*: FUZZY MODEL USED AND HMI VIEW AT THE END OF THE SIMULATION**

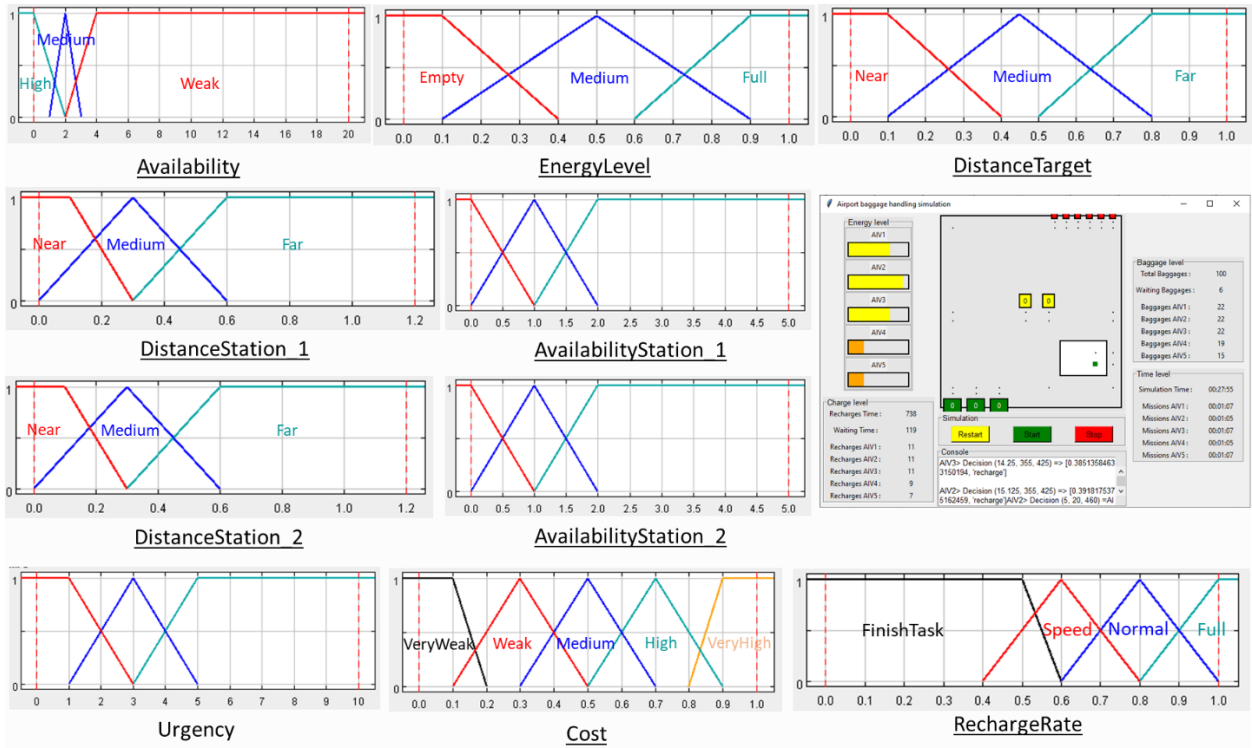

## 5. CONCLUSION

To test various work allocation management scenarios for mobile baggage conveyor robots in the context of Airport 4.0, we developed a multi-agent simulation platform. This methodology provides an adaptable solution to the various facets of AIV autonomy management and enables any necessary modifications for implementation at an airport location. In the event of a central infrastructure failure, the deployment of a distributed multi-agent system offers temporary autonomy and can enhance the control of specific AIV tasks, including task distribution, battery charging, speed regulation, and so on.

We began by developing three fundamental scenarios implementing random, FIFO, and AIV availability techniques in order to provide a foundation for comparing auction-based job allocation strategies with the fuzzy approach we wished to design. Following a test of a task allocation scenario using a simple fuzzy model, we extended the fuzzy decision model of the AIV to include the following features: (1) charging the AIVs' batteries; (2) identifying the recharging station; (3) identifying the most pertinent recharging rate; and (4) controlling the AIVs' speed to allow them to adjust to variations in the baggage arrival flow.

According to the simulation results, the operational efficiency of the AIV fleet can be increased by using adaptive fuzzy multi-agent models for task distribution, energy recharging management, identifying the most advantageous infrastructure components (charging stations), and speed adaptation. These findings demonstrate the value of adaptable and cooperative strategies for enhancing autonomous systems' performance in changing settings.

We plan to continue integrating fuzzy models into AIV agent behavior simulations (Grosset, et al., 2024) and to add learning capabilities (e.g., based on neural networks (Yudha, et al., 2019)) to them in order to increase the relevance and efficiency of their decisions in the collective management of their autonomies.

## ACKNOWLEDGMENTS

The authors would like to thank the Brittany region for funding the *VIASIC* and *ALPHA* projects, as part respectively of the ARED-2021-2024 call for projects, and the PME 2022 call for projects entitled "Accelerate time to market of digital technological innovations from SMEs in the Greater West".